\documentstyle[epsf,preprint,aps]{revtex}
\draft
\preprint{SNUTP 00/023}
\begin{document}
\title{\Large\bf The $\mu$ and soft masses from the intermediate 
scale brane with non-factorizable geometry} 
\author{Jihn E. Kim\footnote{jekim@phyp.snu.ac.kr}
and Bumseok Kyae\footnote{kyae@fire.snu.ac.kr}} 
\address{ Department of Physics and Center for Theoretical
Physics, Seoul National University,
Seoul 151-742, Korea}
\maketitle
\begin{abstract}
The proton decay problem and the negative brane tension problem
in the original Randall-Sundrum model can be resolved by interpreting
the Planck scale brane as the visible sector brane. The hierarchy
problem is resolved with supersymmetry, and the TeV scales for
soft masses and $\mu$ in supersymmetric models are generated by 
the physics at the intermediate scale ($\sim 10^{11-13}$~GeV) brane. 
\end{abstract}

\newpage
 
One of the major theoretical puzzles in particle 
physics is the gauge hierarchy 
probelm~\cite{ghp}, which is basically the Higgs boson mass
parameter(TeV scale) problem in the presence of the
fundamental scale of order the Planck mass($M=2.44\times 10^{18}$ GeV).
The well-known techni-color and supersymmetric solutions 
aim toward obtaining a TeV scale scalar mass naturally~\cite{ghpsol}.
Recently, an alternative solution toward the gauge hierarchy
problem has been suggested by Randall and Sundrum (RS)~\cite{rs}, 
where the huge gap between the Planck and TeV scales is explained 
by the exponetial warp factor of a 5 dimensional space-time metric,   
\begin{equation}
ds^2=e^{-2\sigma(y)}\eta_{\mu\nu}dx^{\mu}dx^{\nu}+b_0^2dy^2~,
\end{equation} 
where $y$ is the fifth dimension coordinate and $\sigma(y)=kb_0|y|$ is given 
as a solution of the Einstein equation. 
For the purpose of obtaining such a metric, Randall and
Sundrum assumed a negative bulk 
cosmological constant $\Lambda_b\equiv -6k^2M^3$, 
and required a $S^1/Z_2$ symmetry in the extradimesion $y$.  
Then the metric becomes non-factorizable.  
In this model, there are two branes (orbifold fixed points) 
as 4 dimensional boundaries,  
Brane 1 (B1) with a positive cosmological constanat $\Lambda_1\equiv 6k_1M^3$ 
at $y=0$, and Brane 2 (B2) with a negative cosmological constant 
$\Lambda_2\equiv 6k_2M^3$ at $y=y_c$.  
Although the RS setup introduces cosmological constants, their solution is 
still static because of the fine-tuning between the bulk and brane cosmological 
constants, $k=k_1=-k_2$, which is a consistency condition in the model.  

On the branes, there could exist `brane fields' that 
live only in the concerned brane.  They correspond to the 
twisted sector fields in string theory, which 
are necessary for anomaly freedom of the 
theory after orbifold compactification \cite{orbifold}.  
Brane fields on B2 are then governed 
by $M_{Pl}e^{-\sigma (y=y_c)}\approx$ TeV scale physics,  
while bulk fields like the graviton and brane fields on B1 are 
governed by Planck scale physics \cite{rs}.   
 
Even though the two scales are easily understood under 
the RS setup, it still has a few problems:

(i) Late cosmology demands that the visible brane is
better to have a positive cosmological constant 
\cite{cline}, which corresponds to B1 in the RS model.  
Since it is inconsistent with the RS's original motivation
that the TeV brane(B2) is the visible brane,   
there have been several propositions \cite{kkl,kkl2,cosmo} such
that B2 has a positive cosmological constant.

(ii) Under the RS setup it is difficult to rescue 
the GUT idea because the cutoff scale at B2 is TeV.  
It is well-known that the GUT scale should be around $3\times 
10^{16}$ GeV with the particle content of
the minimal supersymmetric standard model(MSSM).

(iii) In addition, the RS model has the proton stability problem 
since the relevant interaction scale is expected 
to be TeV.  For $\tau_p>10^{33}$ years, one has to forbid proton decay 
operators up to dimension 14 \cite{kkl,kkl2}, which is a difficult
problem even though it may be achievable in contrived models.

To circumvent the above difficulties within the non-factorizable
geometry, we propose to regard {\it the Planck brane (B1) 
as the visible brane}, in which the standard model (SM) particles live, 
and obtain {\it the TeV scale parameter(s) from a source at B2}.    
To obtain the TeV scale mass parameter(s) naturally, it would be desirable 
to {\it forbid the required TeV mass parameter(s) at B1.} The role of
B2 is to generate the source for TeV scale masses of B1. The effects
of B2 is transmitted to B1 by bulk fields.

For this purpose, it would be necessary to introduce 
an additional symmetry that forbid
unwanted mass parameter(s) at B1, and 
a bulk field as a messenger, which may couple to both brane fields.  
In our simple model, we will assume a global $U(1)_{A}$ symmetry and
enforce the messenger bulk field to carry the $U(1)_A$ quantum number.  
This symmetry will be broken spontaneously due to an interaction with brane 
fields at B2 which plays the role of symmetry breaking source.  
Then the messenger bulk field will get vacuum expectation value(VEV) to give 
the Higgs a TeV scale mass parameter.     

Even if we get a TeV scale at tree level, 
in order to guarantee the stability between the Planck and TeV 
scales, we introduce supersymmetry(SUSY).   
Then, the scale problem in the theory becomes the $\mu$ problem, which is 
the scale problem in supergravity (SUGRA) models~\cite{mu}.  
With $\mu=0$, there exists the Peccei-Quinn(PQ) symmetry~\cite{pq}. 
Therefore, the electroweak symmetry breaking would introduce the unwanted
Peccei-Quinn-Weinberg-Wilczek axion~\cite{pqww}. 
With the $\mu$ term the PQ symmetry is explicitly broken, and there
does not appear the unwanted axion in the model.  
In our case, however, a TeV scale axion decay constant could appear 
if we generate the $\mu$ term at a TeV scale brane. Therefore,
we do not introduce a TeV scale brane at all. Instead, we introduce
an intermediate scale brane B2. This intermediate scale is of
order $10^{11-13}$~GeV. The TeV scale at B1 is generated  
by the bulk field(s) coupled to Higgs doublets 
{\it through non-renormalizable interaction} in the model
described below.  

The low energy SUGRA models also have another scale, the soft
mass parameters which are expected to be around TeV scale. These
soft masses are generated once SUSY is broken. 
The popular SUGRA models use the ideas of hidden sector gaugino 
condensation at intermediate scale, gauge mediated SUSY
breaking, or SUSY breaking due to anomaly.
In our case, we employ the gaugino condensation {\it at B2}, which seems
to be the simplest method.  
At B2, the natural mass scales are 
around the intermediate scale due to the warp factor. If the
$\beta$ function of a nonabelian gauge group is large and negative,
this gauge group confines immediately below the cutoff scale
of B2 and the corresponding gauginos condense around the intermediate mass
scale. In string inspired $E_8\times E_8^\prime$ models, 
the extra factor group $E_8^\prime$
is suitable for this purpose. In supergravity, gravitino mass
is a barometer for the strength of SUSY breaking. 
In our case, the gravitino mass would be of order TeV scale,  
\begin{equation}
m_{3/2}\approx \frac{\langle \lambda\lambda\rangle}{M_P^2}\approx TeV~, 
\end{equation}
for the gaugino condensation scale of order $10^{13}$ GeV.
The soft parameters are also of order TeV scale since 
{\it the gravitino is a bulk field}. We will discuss it again later.   
Thus, we can show that TeV scale SUSY breaking as well as 
TeV scale $\mu$ term can be realized easily in the RS setup 
{\it even if we identify B1 as the visible brane}.  

In this paper, we neglect the backreaction by the brane and bulk fields 
to the background geometry, and we do not discuss the cosmological constant 
problem, which is assumed to be zero by the fine-tuning in the RS model.
The cosmological constants on B1 and B2 are non-zeros.  
However, we will ignore them also because there exists a massless mode of 
the graviton, which gives the {\it flat space action} effectively 
after integrating over the extradimension \cite{kkl2,rs2}. 
This situation does not change in the supersymmetric
generalization of the RS model~\cite{ab} the method of which
is used in this paper.\footnote{
Alternative supersymmetrizations of the RS model in the framework of 
5D supergravity are suggested in \cite{lalak}.}

Let us suppose the global supersymmetry for a simple analysis
and introduce three siglet fields with a global symmetry 
$U(1)_{A}$ which plays the role of the PQ symmetry. The 
$U(1)_{A}$ charges, $A$, of the various fields are shown in Table 1.  

\vskip 0.2cm
\begin{center}
\begin{tabular}{|c||c|c|c|c|c|} \hline
&\ Bulk\ &\multicolumn{2}{|c|}{B1}&\multicolumn{2}{c|}{B2} \\ \hline 
\ Fields\ &$\Sigma^i$&$H_1$, $H_2$&MSSM&$S$&$Z$ \\ \hline 
$A$&$+1$&$-1$&$\frac{1}{2}$&$-1$&$~0~$ \\ \hline 
\end{tabular}
\vskip 0.2cm
{\bf Table 1.~} The $U(1)_A$ charges, $A$ of the various fields 
in the bulk and branes.
\end{center}

\noindent
All the MSSM particles are B1 brane fields, $\Sigma^{i}$ live(s) in the 
bulk and $S$ and $Z$ are B2 brane fields. Of course, the gravity
multiplet which is not shown in Table 1 is of course the bulk field.
Brane fields $S\equiv\left(\phi_S,\psi_S\right)$ and 
$Z\equiv\left(\phi_Z,\psi_Z\right)$ form $D=4$ chiral superfields
at B2. Bulk fields 
$\Sigma^i\equiv \left(\Phi^i, \Psi_{(L,R)}\right)$ constitute $D=5$ 
hypermultiplet, where $\Phi^i$ $(i=1,2)$ are two complex scalars and 
$\Psi_{(L,R)}$ are left and right handed Dirac fermions.   
We define $\Psi_{L,R}$ as $\gamma_5\Psi_{L,R}=\pm\Psi_{L,R}$.  
They constitute a $N=2$ supermultiplet and make the theory non-chiral.   

At B1, all dimension 2 and 3 operators containing $H_1H_2$, 
including the so-called $\mu$ term, 
are forbidden by the $U(1)_{A}$ symmetry. The dominant term
in the superpotential consistent with the $U(1)_A$
symmetry is the dimension 4 operator,  
\begin{equation}
\sim \frac{\Sigma^2}{M_{P}}H_1H_2~.  
\end{equation}
This is just a {\it schematic formula}; 
it should be rewritten such that {\it the interactions between the 
bulk and brane fields respect $D=4$ SUSY after integrating over the 
extradimension}, which is described below.    
At B2, the most general dimension 3 superpotential is 
\begin{equation}
\sim Z\left(\Sigma S-M_{P}^2\right)~, 
\end{equation} 
which is also a {\it schematic formula}.  

The action of the hypermultiplet in the bulk is     
\begin{eqnarray} \label{action}
S_{bulk}=-\sum_{i}\int d^5x \sqrt{-G}\bigg[
g^{MN}&&\partial_M\Phi^{i*}\partial_N\Phi^{i}
+\frac{i}{2}\left(\overline{\Psi}\Gamma^M\nabla_M\Psi
-(\nabla_M\overline{\Psi})\Gamma^M\Psi\right) \nonumber \\
&&+M_{\Phi^{i}}^2|\Phi^{i}|^2+M_{\Psi}\overline{\Psi_{L}}\Psi_{R}
+M_{\Psi}\overline{\Psi_{R}}\Psi_{L}\bigg]~~,
\end{eqnarray}
where $\Gamma^M\equiv e^M_a\gamma^a$.  
$\nabla_M$ is the covariant derivative on a curved manifold and defined as 
$\nabla_M\equiv \partial_M+\omega_M$, where 
$\omega_M$ is the spin connection,  
\begin{equation}
\omega_\mu=\frac{1}{2}\gamma_5\gamma_\mu\frac{d\sigma(y)}{dy} 
~~~{\rm and}~~~ \omega_5=0~~~.  
\end{equation}  
Invariance under the supersymmetry transformations \cite{shuster,gp},
\begin{eqnarray}
\delta \Phi^{i}&=&i\sqrt{2}\epsilon^{ij}\bar{\eta^j}\Psi \\
\delta \Psi&=&\sqrt{2}\left[\Gamma^M\partial_M\Phi^{i}\epsilon^{ij}
-\frac{3}{2}\sigma'\Phi^{i}(\epsilon\sigma_3)^{ij}
-M_{\Psi}\Phi^{i}\epsilon^{ij}\right]\eta^j~~,
\end{eqnarray}
requires that the five dimensional masses of the scalars and fermions satisfy 
\begin{eqnarray}
M_{\Phi^{1}}^2&=&(t^2+t-\frac{15}{4})\sigma^{'2}+(\frac{3}{2}-t)\sigma'' 
\label{mh1}\\
M_{\Phi^{2}}^2&=&(t^2-t-\frac{15}{4})\sigma^{'2}+(\frac{3}{2}+t)\sigma'' \\ 
M_{\Psi}&=&t\sigma'~~,
\end{eqnarray}
where $\sigma'\equiv kb_0\bigg[2\bigg(\theta(y)-\theta(y-y_c)\bigg)-1\bigg]$ and 
$\sigma''\equiv 2kb_0\bigg(\delta(y)-\delta(y-y_c)\bigg)$, 
and $t$ is an arbitrary dimensionless parameter.  
In flat space-time, supersymmetry requires the same masses for the scalars and 
fermions.  But in the 
$AdS_5$ background, the fields in the same supermultiplet 
must have different masses \cite{shuster,gp}.  

Equations of motion for the above action allow the massless modes \cite{gp},     
\begin{eqnarray}
\Phi^{1,(0)}(x,y)&=&\frac{e^{(3/2-t)\sigma(y)}}{\sqrt{2b_0y_c}N}\phi_{\Sigma}(x)
\label{tphi}\\
\Psi_{L}^{(0)}(x,y)&=&\frac{e^{(2-t)\sigma(y)}}{\sqrt{2b_0y_c}N}\psi_{\Sigma}(x)
~~, \label{tpsi}
\end{eqnarray}
where $\sigma(y)\equiv kb_0|y|$ and the normalization factor $N$ is given by
\begin{equation} \label{norm}
N^2\equiv\frac{e^{\sigma_c(1-2t)}-1}{\sigma_c(1-2t)}~~,  
\end{equation}
where $\sigma_c\equiv kb_0y_c$ and $N$ is reduced to 1 when $t=1/2$.    
Note that {\it the massless modes depend on `$y$'}.  
$\Phi^{2}(x,y)$ and $\Psi_{R}(x,y)$ do not have massless modes 
since they are inconsistent with the orbifold condition \cite{gp}, and hence 
the fermion mass terms cannot exist for the massless modes.  
We will see below that the scalar mass terms are canceled also 
in the effective 4 dimensional action.  
When decoupling the Kaluza-Klein massive modes, 
$N=2$ SUSY breaks down to $N=1$ SUSY 
for the massless modes of the hypermultiplet, 
and $\Sigma^{1,(0)}\equiv\left(\Phi^{1,(0)}(x,y),\Psi_{L}^{(0)}(x,y)\right)$ 
form $N=1$ supersymmetric chiral multiplets.  

Inserting Eqs.~(\ref{mh1})--(\ref{norm}) to Eq.~(\ref{action}) 
and integrating by parts, we get the 4 dimensional effective 
action for the massless modes,     
\begin{eqnarray}
S_{bulk}^{eff}&=&-\int d^4x\int_{-y_c}^{y_c}dy 
\bigg[\frac{b_0e^{-4\sigma}}{2b_0y_cN^2}
\bigg(e^{2\sigma}e^{(3-2t)\sigma}\eta^{\mu\nu}
\partial_{\mu}\phi_{\Sigma}^{*}\partial_{\nu}\phi_{\Sigma}
+e^{(4-2t)\sigma}e^{\sigma}
i\overline{\psi_{\Sigma}}\gamma^{\mu}\partial_{\mu}\psi_{\Sigma}\bigg)
\nonumber\\
~~~&&+\frac{b_0}{2b_0y_cN^2}\left(-e^{(\frac{3}{2}-t)\sigma}\partial_y(
e^{-4\sigma}\partial_ye^{(\frac{3}{2}-t)\sigma})|\phi_{\Sigma}|^2
+e^{-4\sigma}M_{\Phi^1}^2e^{(3-2t)\sigma}|\phi_{\Sigma}|^2\right)\bigg]
\nonumber\\
&=&-\left(\frac{1}{2y_cN^2}\int_{-y_c}^{y_c}e^{(1-2t)\sigma(y)}\right)
\int d^4x\bigg[
\eta^{\mu\nu}\partial_{\mu}\phi_{\Sigma}^{*}\partial_{\nu}\phi_{\Sigma}
+i\overline{\psi_{\Sigma}}\gamma^{\mu}\partial_{\mu}\psi_{\Sigma}\bigg]
\nonumber\\ 
&=&-\int d^4x\bigg[
\eta^{\mu\nu}\partial_{\mu}\phi_{\Sigma}^{*}\partial_{\nu}\phi_{\Sigma}
+i\overline{\psi_{\Sigma}}\gamma^{\mu}\partial_{\mu}\psi_{\Sigma}\bigg]~,  
\end{eqnarray}
where we see the scalar mass terms are eliminated and 
the contributions of $\partial_y\Psi$ and $\partial_y\overline{\Psi}$
add up to zero.
Thus, we confirm that the $\phi_{\Sigma}$ and $\psi_{\Sigma}$ are 
massless fields. In the above equations, $t$ is not fixed yet.  
However, if the bulk fields are required to couple {\it supersymmetrically} 
to the brane fields on a brane, $t$ should be fixed to $\frac{1}{2}$ 
and the bulk fields form a 4 dimensional supermultiplet as 
\begin{equation}\label{4dbulk}
\bar{\Sigma}(x,y)=\left(e^{\sigma(y)}\phi_{\Sigma}(x),
e^{\frac{3}{2}\sigma(y)}\psi_{\Sigma}(x)\right)~~,   
\end{equation}
which will become clear below.  

At the intermediate brane B2, the brane 
fields are required to be rescaled 
such that their kinetic terms have the canonical forms, 
\begin{eqnarray}
S_{B2}^{kin}
&=&-\sum_{i=S,Z}\int d^4xe^{-4\sigma_c}\bigg[e^{2\sigma_c}\eta^{\mu\nu}
\partial_{\mu}\phi_{i}^{*}\partial_{\nu}\phi_{i}
+e^{\sigma_c}i\overline{\psi_{i}}\gamma^{\mu}\partial_{\mu}\psi_{i}\bigg]
\nonumber \\ 
&=&-\sum_{i=S,Z}\int d^4x\bigg[\eta^{\mu\nu}
\partial_{\mu}\tilde{\phi}_{i}^{*}\partial_{\nu}\tilde{\phi}_{i}
+i\overline{\tilde{\psi}_{i}}\gamma^{\mu}\partial_{\mu}\tilde{\psi}_{i}\bigg]~~,
\end{eqnarray}
where we use $e^{a}\,_{\mu}|_{y=y_c}=e^{-\sigma_c}\delta^{a}_{\mu}$ 
and $\tilde{\phi}_{i}$ and $\tilde{\psi}_{i}$ are defined 
as the rescaled fields as follows, 
\begin{equation}\label{irescale}
\phi_{i}\equiv e^{\sigma_c}\tilde{\phi}_{i}~~~{\rm and}~~~
\psi_{i}\equiv e^{\frac{3}{2}\sigma_c}\tilde{\psi}_{i}~~~.
\end{equation}

At B2, the brane fields $S=\left(\phi_S,\psi_S\right)$ and 
$Z=\left(\phi_Z,\psi_Z\right)$, and bulk fields 
$\bar{\Sigma}^{1,(0)}|_{y=y_c}=\left(e^{\sigma_c}\phi_{\Sigma}(x),
e^{\frac{3}{2}\sigma_c}\psi_{\Sigma}(x)\right)$ can form Yukawa interaction 
terms {\it supersymmetrically} in the action, 
\begin{eqnarray}\label{iint}
S_{B2}^{int}&=&\int d^4x\sqrt{-g_4}~\bigg[
\left((e^{\sigma_c}\phi_{\Sigma})\psi_S\psi_Z
+\phi_S\psi_Z(e^{\frac{3}{2}\sigma_c}\psi_\Sigma)
+\phi_Z(e^{\frac{3}{2}\sigma_c}\psi_\Sigma)\psi_S+{\rm h.c.}\right)\nonumber \\
~~~~~&&-|\phi_S\phi_Z|^2-|\phi_Z(e^{\sigma_c}\phi_{\Sigma})|^2
-|(e^{\sigma_c}\phi_{\Sigma})\phi_S|^2
+M_P^2\left((e^{\sigma_c}\phi_{\Sigma})\phi_S+{\rm h.c.}\right)-M_P^4\bigg]
\nonumber\\
&=&\int d^4x~\bigg[\left(\phi_{\Sigma}\tilde{\psi}_S\tilde{\psi}_Z
+\tilde{\phi}_S\tilde{\psi}_Z\psi_\Sigma
+\tilde{\phi}_Z\psi_\Sigma\tilde{\psi}_S+{\rm h.c.}\right) \nonumber \\ 
~~~~~&&-|\tilde{\phi}_S\tilde{\phi}_Z|^2-|\tilde{\phi_Z}\phi_{\Sigma}|^2
-|\phi_{\Sigma}\tilde{\phi}_S-m_I^2|^2
\bigg]~~,
\end{eqnarray}
where we use $\sqrt{-g_4}~|_{y=y_c}=e^{-4\sigma_c}$ 
and Eq.~(\ref{irescale}).  
$m_I$ is defined as $m_I\equiv M_Pe^{-\sigma_c}
\sim 10^{11}-10^{13}$ GeV 
for $\sigma_c=kb_0y_c\approx 11.5-16$.  
Here we use the Weyl spinor notation for spinor fields.  
Eq.~(\ref{iint}) is clearly invariant under the $D=4$ SUSY transformations.  
To see that explicitly, let us define 
\begin{equation}
\tilde{\Sigma}(x)\equiv 
\Bigg(\phi_{\Sigma}(x),\psi_{\Sigma}(x)\Bigg),~~~ 
\tilde{S}(x)\equiv\Bigg(\tilde{\phi}_S(x),
\tilde{\psi}_S(x)\Bigg)~~~{\rm and}~~~  
\tilde{Z}(x)\equiv\Bigg(\tilde{\phi}_Z(x),\tilde{\psi}_Z(x)\Bigg). 
\end{equation}
Then we can derive 
a superpotential at B2 from Eq.~(\ref{iint}) as follows, 
\begin{equation} \label{wb2}
W_{B2}=\tilde{Z}\left(\tilde{\Sigma} \tilde{S}-m_I^2\right)~~~.
\end{equation} 
Here we note that {\it if we did not choose 
$t=\frac{1}{2}$ in Eqs.~(\ref{tphi}) 
and (\ref{tpsi}), we could not get the $D=4$ effective supersymmetric 
interactions between bulk and brane fields}.  

At B1 all the needed Yukawa interactions in the 
MSSM except the $\mu$ term are allowed. The nonrenormalizable
interactions are expected in the supergravity generalization of
our global SUSY study. In this case, we may define
the superpotential $W$ as the maximum~\cite{kim99} 
allowed by the symmetry
\begin{equation}
G=K+M_P^2\log \frac{|W|^2}{M_P^6}={\rm invariant}
\end{equation}   
where $K$ is a K$\ddot{\rm a}$hler potential. In this case,
we generally introduce all possible non-renormalizable terms
consistent with the symmetry. Typically, the following superpotential
is present, 
\begin{equation} \label{wb1}
W_{B1}=\frac{\tilde{\Sigma}^2}{M_P}H_1H_2~.
\end{equation}
$^{}$From Eqs.~(\ref{wb2}) and (\ref{wb1}), 
we derive supersymmetric scalar potential, 
\begin{eqnarray}
V&=&V_{B1}+V_{B2}
\simeq\sum_{p=\Sigma,H_1,H_2}
|\frac{\partial W_{B1}}{\partial \phi_p}|^2
+\sum_{i=\Sigma,\tilde{S},\tilde{Z}}
|\frac{\partial W_{B2}}{\partial \phi_i}|^2
\nonumber \\
&=&|\frac{\phi_{\Sigma}^2}{M_P}\phi_{H_1}|^2
+|\frac{\phi_{\Sigma}^2}{M_P}\phi_{H_2}|^2
+|\frac{2\phi_{\Sigma}}{M_P}\phi_{H_1}\phi_{H_2}|^2
+|\tilde{\phi}_S\tilde{\phi}_Z|^2
+|\tilde{\phi}_Z\phi_{\Sigma}|^2+|\phi_{\Sigma}\tilde{\phi}_S-m_I^2|^2~.
\end{eqnarray} 
The potential has a minimum at $\phi_{\Sigma}\tilde{\phi}_S=m_I^2$ and 
$\tilde{\phi}_Z=\phi_{H_1}=\phi_{H_2}=0$.  
But TeV scale SUSY breaking would lift this degeneracy as much as TeV scale.  
Thus $\phi_Z$, $\phi_{H_1}$ and $\phi_{H_2}$ could have VEV of TeV scale,
and the electroweak mass scales are derived. The PQ symmetry breaking
scale is at the intermediate scale, and there results a very light
axion~\cite{light}.  

Since $\phi_{\Sigma}$ obtains an intermediate scale VEV
\footnote{In a model such that the $\mu$ term is generated 
from $W_{B1}=\frac{\langle\phi_{\Sigma}\rangle^3}{M_P^2}H_1H_2$,  
$10^{13}$ GeV would be preferred as an intermediate scale.  }, 
we can get a TeV scale $\mu$ term at B1, 
\begin{equation}
W_{B1}\approx \frac{\langle\phi_{\Sigma}\rangle^2}{M_P}H_1H_2~. 
\end{equation}     
This form of the $\mu$ term from the symmetry principle
has been considered before~\cite{mu,ckn,kim99}. Here,
we realize it with the intermediate scale brane world.

Similarly, we expect a TeV scale gravitino mass. Let us
suppose that the gravitino mass is generated at $B2$. At 
the Planck brane $B1$
a bilinear scalar field $\phi^*\phi$ couples to two gravitino lines
with a coupling suppressed by the Planck mass $M_{Pl}$. 
These gravitinos propagate in the bulk
and at $B2$ they meet to produce the gravitino mass. Since the bulk
propagator cutoff ranges up to $M_{Pl}$, the resulting effective gravitino
mass after the Feynman integration would be of order TeV scale
since the gravitino mass at $B2$ is of order TeV scale. 
But without the gravitino, the soft mass would
not be generated. For example, the graviton coupling to $\phi^*\phi$
at $B1$ is through a derivative of the scalar field, which means that
for the vanishing momentum of $\phi$ the bulk propagation of the graviton
would give a vanishing potential and hence no soft mass contribution.
Namely, the SUGRA is needed to obtain the soft mass terms
of the scalars living at $B1$.
For the gravitino propagation calculation, however, we need
an exact expression for the gravitino propagator in the bulk. Even though
we have that expression, we must cutoff at the Planck scale for the
above gravitino loop and the estimate is anyway a kind of order of
magnitude. 
    
Therefore, it is better and clear to use an effective 4 dimensional
theory after integrating the $y$ coordinate. Here, we will use the 
SUGRA language. In order to trigger the SUSY breaking with the gaugino 
condensation, let us introduce {\it on B2} a vector multiplet(corresponding
to the confining hidden sector force), 
$(A^{a}_{\mu}(x), \chi^a \equiv e^{\frac{3}{2}\sigma_c}\lambda^a (x))$.
We can easily check that $A^a_{\mu}(x)$ and $\lambda^a(x)$ give the canonical 
kinetic terms through the above procedure, Eq.~(17)--(\ref{irescale}). 

The graviton and gravitino, which are bulk fields, have the massless modes 
$(e^a\,_{\mu}(x,y)\sim e^{-\sigma(y)}\tilde{e}^a\,_{\mu}(x); 
\Psi^{(1,2)}_{\mu}(x,y)\sim e^{-\frac{1}{2}\sigma(y)}\psi_{\mu}(x))$ 
\cite{ab,lalak2,rs2,kkl2}.  
Thus, after decoupling the KK modes and integrating over the extra
dimension and following the above procedure Eq.~(\ref{action})--(15), 
we can confirm that the $e^a_{\mu}$ and $\psi_{\mu}$ are the effective
4 dimensional graviton and gravitino, which are actually massless 
at least whithout any other physics \cite{ab}, 
\begin{eqnarray}
S^{kin}_{bulk}&=&\int d^4x\int_{-y_c}^{y_c}dy\bigg[
\sqrt{-g}\frac{1}{2}R-\Lambda_b +(\frac{-\Lambda_b}{6})\epsilon^{MNOPQ}
\overline{\Psi}_O\Sigma_{PQ}D_Q\Psi_N +\dots\bigg]\nonumber \\
&=&\int d^4x\bigg[\sqrt{-\tilde{g_4}}\frac{M^2_{Pl}}{2}\tilde{R}_4
+\epsilon^{\mu \nu \rho \sigma}\overline{\psi}_{\mu}\overline{\sigma}_{\nu}
D_{\rho}\psi_{\sigma}\bigg], 
\end{eqnarray}
where we set $M$ as 1 and use the RS fine-tuning conditions between 
the bulk and brane cosmological constants, 
$\sqrt{-\Lambda_b/6}=\Lambda_1/6=-\Lambda_2/6$.  
We note here that {\it the effective 4 dimensional space-time is flat}.     

After decoupling the KK modes, however, 
the massless mode of gravitino and the {\it brane} gaugino field 
could compose the 4-fermion interaction terms at B2 in the supersymmetric way
\cite{hm,lalak2},  
\begin{eqnarray}
S^{4-fermi}_{B2}&\sim&\int d^4x \sqrt{-g_4}\bigg[
\frac{e^{3\sigma_c}\langle\lambda^a\lambda^a\rangle}{M^2_P}
e^{-\frac{1}{2}\sigma_c}\overline{\psi}_{\mu}
\Sigma^{\mu \nu}e^{-\frac{1}{2}\sigma_c}\psi_{\nu}+\dots\bigg] \nonumber \\
&=&\int d^4x\bigg[\frac{\langle\lambda^a\lambda^a\rangle}
{M^2_P}\overline{\psi}_{\mu}\sigma^{\mu \nu}\psi_{\nu}+\dots\bigg]
\end{eqnarray}
where we use $e^a\,_{\mu}|_{y=y_c}=e^{-\sigma_c}\delta^a_{\mu}$ and 
$\Sigma^{\mu}\equiv e^{\mu}\,_a\sigma^a=(e^{\sigma_c}\delta^{\mu}_a)\sigma^a$.  
Thus the gravitino mass term, 
which could be a barometer for SUSY breaking, 
is generated in our effective 4 dimensional theory  
when the brane gauginos condense. Since the intermediate scale 
($\sim 10^{13}$GeV) is a natural cutoff and the 
condensation scale on B2 {\it with a large gauge group},  
the above term gives a TeV scale gravitino mass successfully, 
as expained before.  

After integraing over the extra dimension, 
the hidden brane loses its geometrical meaning    
and just remains as a {\it hidden sector gauge field}. 
Thus we obtain an effective 4 dimensional theory again 
{\it with a visible gauge sector whose tipical cutoff scale is the 
Planck scale, and a hidden gauge sector whose cutoff scale is the 
intermediate scale}.  
The 4 dimensional gravity sector fields, which was in the bulk before, 
interact with all the fields irrespective of which sector they come from. 
Therefore, in the effective theory language it is {\it exactly 
the same picture with the 
conventional D=4, N=1 SUGRA scenario with the hidden sector, except 
that the hidden sector's cutoff scale is the intermediate scale}.  
Thus, the soft mass terms of the scalar fields are the square of 
the gravitino mass $m_{3/2}$ which is the same as the
$B_2$ gravitino mass given in Eq.~(27) since the $y$ integration
for $m_{3/2}$ gets a contribution only from $B2$ through the Dirac
delta function $\delta(y-y_c)$.

The effective 4 dimensional theory gives possibly a consistent 
D=4, N=1 SUGRA. In this case, all kinds of SUSY breaking coefficients 
in the Lagrangian are parameterized with the gravitino 
mass.  Therefore, we could get the TeV scale SUSY breaking effects 
from the TeV scale gravitino mass through the gravitational 
interaction in the bulk.      
    
In conclusion, we constructed a successful brane model
where the visible sector lives in the Planck scale brane,
and the supersymmetry breaking occurs at the intermediate scale
brane. A TeV brane is not needed. The TeV scale, the soft
masses and $\mu$, are derived
quantities from the mass source at the intermediate scale brane,
through the mediation by bulk fields. The proton decay problem
is resolved since the grand unification is achieved at the
Planck brane. Since the visible sector is at B1, the problem
of negative brane tension in the original RS model does not arise.
For the stability of soft masses and $\mu$, we introduced
supersymmetry, which is possible for a specific value of 
the parameter $t$. 
 
\acknowledgments
This work is supported in part by the BK21 program of Ministry 
of Education, Korea Research Foundation Grant No. KRF-2000-015-DP0072, 
CTP Research Fund of Seoul National University,
and by the Center for High Energy Physics(CHEP),
Kyungpook National University.

\end{document}